\documentclass[%
 reprint,
 amsmath,amssymb,
 aps,
]{revtex4-1}

\usepackage{hyperref}
\usepackage{kpfonts}
\usepackage[T1]{fontenc}

\usepackage{amsmath,amscd}
\usepackage{enumerate}
\usepackage{amsmath,amscd}

\newcommand{\ket}[1]{\left| #1 \right>} 
\newcommand{\bra}[1]{\left< #1 \right|} 
\let\baraccent=\= 
\renewcommand{\=}[1]{\stackrel{#1}{=}} 

\usepackage{mathtools}
\usepackage{tensor}
\usepackage{tikz}
\usepackage[all,cmtip]{xy}

\usetikzlibrary{matrix,arrows,decorations.pathmorphing}

\providecommand{\CC}{\mathbb{C}}

\providecommand{\ZZ}{\mathbb{Z}}
\providecommand{\e}{\epsilon}
\newtheorem{thm}{Theorem}[section]

\usepackage{cancel}
\usepackage{kpfonts}
\usepackage[T1]{fontenc}

\providecommand{\fr}{\frac}

\usepackage{graphicx}
\usepackage{dcolumn}
\usepackage{bm}

\usepackage{graphicx}
\usepackage{dcolumn}
\usepackage{bm}

\begin{document}

\date{ December 18, 2016}

\title{Monopole-Enriched Groundstates of Two-Dimensional BCS Models}

\thanks{Honors thesis written under the supervision of Clifford Taubes}%

\author{David Roberts$^{1,2}$}
\affiliation{%
$^1$Department of Physics, Harvard University, Cambridge MA 02138, USA\\
$^2$NASA Quantum Artificial Intelligence Laboratory, Moffett Field, CA 94035, USA
}%

\begin{abstract}  We extract the macroscopic characteristics of the groundstate sectors of two dimensional Bose-condensed BCS models with a fixed  number of magnetic flux quanta, on length scales much larger than the characteristic size of the Cooper-pair bound states. We show that this reduces to the problem of computing the moduli space of gauge-inequivalent solutions to the Gross-Pitaevskii equations on a nontrivial fibre bundle. Inspired in part by the physical arguments of Oshikawa and Senthil in {\it Fractionalization, Topological Order, and Quasiparticle Statistics}, we extract a large class of groundstates from this moduli space.\end{abstract}

\maketitle

\section{Introduction}
One of the hallmarks of a topological phase of matter is a marked dependence of the number of groundstates on the topology of the system. Superconductors share this property, in that the groundstate degeneracy of a superconductor shaped like a manifold $\Sigma$ is a topological invariant of $\Sigma$. However, topological groundstate degeneracy in superconductors with non-zero net magnetic flux is less well-understood.  Therefore, in this paper, we address this gap and compute the groundstates of two-dimensional superconductors with background monopoles. We find a continuum of superconducting groundstates, as well as other exotic physics. \\

Our model of the superconductor is the standard BCS model with the $s$-wave pairing hypothesis, in the ultra-dilute, infinite volume limit, in which the BCS energy density on $\Sigma$ becomes well-approximated by the Gross-Pitaevskii energy density. The set of physically-distinct superconducting groundstates is then the moduli space of gauge-inequivalent solutions to the Gross-Pitaevskii equations on a complex line bundle $\mathcal L\to\Sigma$. Therefore, we use the variational principle of quantum mechanics to extract the superconducting groundstates.\\

In this variational method, the difficulty of extracting the superconducting groundstates with monopoles translates into the difficulty of solving the GP equations on a nontrivial line bundle $\mathcal L$.  Despite this difficulty, we surmount  it using powerful mathematical methods, inspired in part by geometric analysis  (Taubes's PhD thesis {\it Vortices and Monopoles}), and in part by condensed matter physics (Oshikawa and Senthil in {\it Fractionalization, Topological Order, and Quasiparticle Statistics}).

\section{~~~~~~Background}
\subsection{~~~~~Quantization of Fermions on a Riemann Surface}
To even discuss the BCS model on a nontrivial surface requires some work. A great way to start is to construct the Hilbert spaces involved, as well as the local field algebra of observables, in a manifestly covariant way. \\
\begin{center}
\includegraphics[scale=.2]{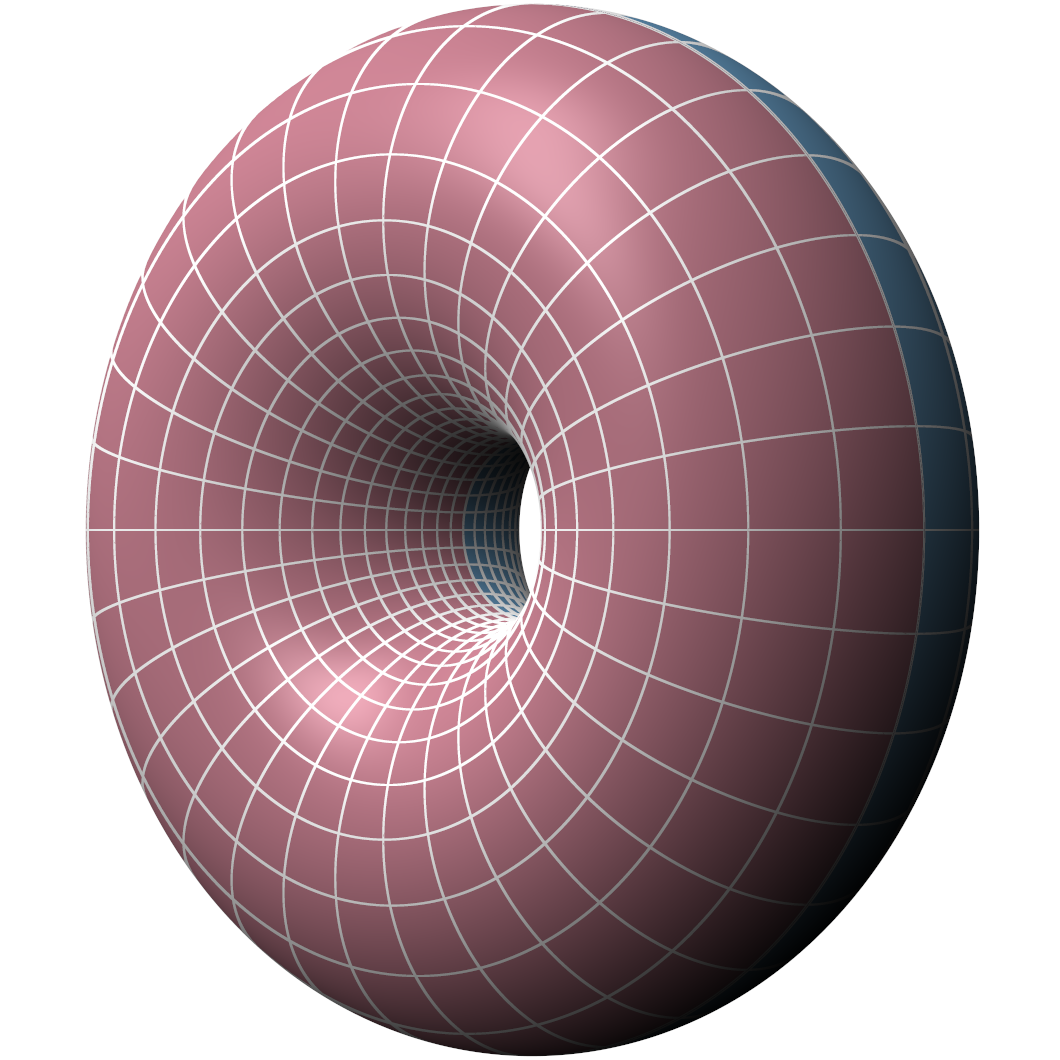}\\
{\small {\bf Figure 1.} The superconductor $\Sigma$. In this case, $g=1$ (Source: Google Images).}
\end{center}
We begin with the observation that the space of states for a fermion on $\Sigma$ is {\it unitarily} isomorphic to the Hilbert-space $L^2(\mathcal L)$ of square-integrable cross-sections of some hermitian line bundle $\mathcal L$ over $\Sigma$.\\

Now, if we fix a surface $\Sigma$, there are many line bundles $\mathcal L\to \Sigma$. In fact a standard result of the theory of characteristic classes is that these are classified modulo diffeomorphisms by their first Chern class. By the Chern-Weil homorphism, this integer is precisely the number of magnetic flux quanta penetrating the surface:
\[c_1(\mathcal L)\in H^2(\Sigma)\simeq \ZZ\]
$n$ is also commonly known as the magnetic monopole number, but this is only a convenient device for physical intuition; there is no "outside" of the surface.\\

We can then ask: what line bundle should we consider for our model? In this paper, we allow all possibilities: each fixed $n$ defines a unique line bundle $\mathcal L$ up to diffeomorphisms. Thus each topological sector contains a distinct smooth structure  and ultimately a full-fledged BCS model. The construction goes like this: having fixed an integer $n$, we then define an electromagnetic gauge field $A$, inducing a holomorphic structure on $\mathcal L$.  In the corresponding Yang-Mills theory, the dynamical variable is the {\it holomorphic structure} of $\mathcal L$: as the gauge field evolves through time, the holomorphic structure of $\mathcal L$ may vary through time. In contrast, the topological structure of $\mathcal L$ is static, i.e. {\bf magnetic charge is conserved}. \\

Having fixed a topological structure, we can then construct the corresponding many-body Hilbert space $\mathfrak H$ for the fermions as the exterior algebra on the Hilbert space of square-integrable cross-sections of $\mathcal L$ (here, $n$ indexes fermion, not monopole, number):
\[\mathfrak H:=\oplus_{n\geq 0}\,\Lambda^n(\mathfrak h\,),\tag{$\mathfrak h=L^2(\mathcal L)\otimes \,\CC^2$}\]
The observable algebra $\mathcal O$ can then also be constructed, as the union of local algebras $\{\mathcal O_U\}$ generated by fermonic creation and annihilation operators of cross-sections supported on open subsets $U$ of $\Sigma$.

\subsection{~~~~~The BCS Energy Density}
This and the next subsection will both closely model the discussions in \hyperlink{ref1}{[1]}. The many-fermion model on $\Sigma$ that we wish to study is a standard BCS model with $s$-wave pairing. 
The groundstate postulates of BCS theory with  $s$-wave pairing can be stated succinctly as
\begin{align*}
\min_{|\psi\rangle\,\in \,\mathfrak H} \langle\psi|H_{BCS}|\psi\rangle=\min_{\varphi_0,\varphi_1}\bra{\varphi_0,\varphi_1}H_{BCS}\ket{\varphi_0,\varphi_1},
\end{align*}
where the $s$-wave {\bf BCS variational ansatz} $\ket{\varphi_0,\varphi_1}$ is defined to be the unique quasi-free state determined by the following two-point functions on the fermonic field algebra $\mathcal O$:
\begin{align*}
\varphi_0(x,y)=\langle \Psi_{\uparrow}^\dag(x)\Psi_{\uparrow}(y)\rangle,~~~~\varphi_1(x,y)=\langle \Psi_{\uparrow}^\dag (x)\Psi_{\downarrow}^\dag (y)\rangle.
\end{align*}
The final relevant assumption of the BCS model with $s$-wave pairing is that the expectation value of the energy of the BCS variational ansatz (i.e. the {\bf BCS energy functional}), is assumed to satisfy (see \hyperlink{ref1}{[1]}):
\begin{align*}
\mathcal E_{BCS}&=\mathcal E_{YM}+\int_\Sigma dx\,\,(\nabla^2+U(x))\varphi_0(x,x) \\
& +\fr{1}{2}\int_{\Sigma\times \Sigma}dx\, dy\,V(x-y)|\varphi_1(x,y)|^2.
\end{align*}
Here, $\mathcal E_{YM}$ denotes the standard Yang-Mills energy density of $\nabla$, which is the covariant derivative on $\mathcal L$ specified by the gauge field (here, it is acting on the second argument of $\varphi_0$). Having defined the energy functional, the variational principle yields the set of groundstates for the BCS model:
\begin{align*}
\mathcal H_{BCS}(\Sigma):=\{(A,\varphi_0,\varphi_1)~\text{s.t.}~\mathcal E_{BCS}~\text{is at a minimum}\}
\end{align*}
However, because there is no additional structure on $\mathcal L$ that allows an observer to distinguish between isometric gauges, we must physically identify two groundstates which are related by a gauge transformation. Mathematically, this corresponds to modding-out $\mathcal H_{BCS}$ by the action of the gauge group.\\

Accordingly, we define the {\bf moduli space} of $s$-wave superconducting groundstates on a surface $\Sigma$ by \[\mathcal M_{BCS}(\Sigma):=\mathcal H_{BCS}(\Sigma)/\mathcal G\]
where $\mathcal G$ denotes the action of gauge transformations on the triple $(A,\varphi_0, \varphi_1)$ induced by gauge transformations of the fermonic field algebra $\mathcal O$. Because we have now modded-out by gauge transformations, $\mathcal M_{BCS}(\Sigma)$ is equal to the set of physically-distinguishable groundstates of a BCS model on $\Sigma$.

%

\subsection{~~~~The Macroscopic Limit and the GP Energy Density}
Following \hyperlink{ref1}{[1]}, we introduce an ultra-dilute infinite-volume limit $\e\to 0$, in which the expected number of fermions goes to zero as $\e$, and the volume of the surface approaches infinity as $1/\e^2$. Based on the results of \hyperlink{ref1}{[1]}, we expect $\mathcal E_{BCS}$ to be well-approximated by the {\bf Gross-Pitaevskii energy functional}
\begin{align*}
\mathcal E_{GP}^\alpha&=\mathcal E_{YM}\\
&~~~~+\int_\Sigma|\nabla^{(2)}\psi(x)|^2\,+4U(x)|\psi(x)|^2+\alpha |\psi(x)|^4
\end{align*}
i.e. the energy density depends only on an order parameter $\psi\in L^2(\mathcal L\otimes \mathcal L)$ for paired electrons. This is expected, because the superconducting state, according to the BCS theory, is a BEC of paired electrons, and the Gross-Pitaevskii functional is well-known to describe the energy of a BEC, as proven by Erdos et. al. in \hyperlink{ref2}{[2]}. Here,
\[\nabla^{(k)}(\psi_1\otimes\cdots \psi_k):= \sum_{j=1}^k \psi_1\otimes\cdots \nabla \psi_j \otimes \cdots \psi_k,\]
i.e., the covariant derivative on the Cooper-pairs is constructed via second-quantization of the single-particle covariant derivative. Therefore, in this ultra-dilute, infinite-volume limit, the set of groundstates for the $s$-wave BCS model is, effectively
\[\mathcal H_{GP}^\alpha(\Sigma):=\{(A,\psi)~\text{s.t.}~\mathcal E_{GP}^\alpha~\text{is at a minimum}\}\]
Again, because there is no structure on $\mathcal L$ that allows an observer to distinguish between isometric gauges, we must mod-out the action of the gauge group on $\mathcal H_{GP}^\alpha$.\\

Accordingly, the {\bf moduli space} of solutions to the GP equations on a surface $\Sigma$ modulo gauge transformations is defined as
 \[\mathcal M_{GP}^\alpha(\Sigma):=\mathcal H_{GP}^\alpha(\Sigma)/\mathcal G\]
where $\mathcal G$ denotes the action of gauge transformations on the pair $(A,\psi)$ induced by gauge transformations of the fermonic field algebra $\mathcal O$. Because we have now modded-out by gauge transformations, $\mathcal M_{GP}(\Sigma)$ is equal to the set of physically-distinguishable groundstates of a BCS model on $\Sigma$, in this ultra-dilute, infinite-volume limit.

\section{~~~~~~The BCS Groundstate Sector without Monopoles}
Here, we consider the case $n=0$, where the analysis will reproduce N. Read and Green's fascinating result in \hyperlink{ref3}{[3]}, namely, that a BCS model on a Riemann surface has ground states exactly corresponding to the spin structures on that surface, without referring to details of the BCS Hamiltonian. All of this power comes at the price that our results are only relevant in the ultra-dilute, infinite-volume limit.\\

To analyze the GP functional in the $n=0$ sector, we break it up into three parts. Also, to simplify the analysis, we will set the single-particle potential $U\in C^\infty(\Sigma)$ to be a constant. In this case, we can write
\begin{align*}
\mathcal E_{GP}&=\sum_{j=1}^3\mathcal E_j,\\
\mathcal E_1&:=\int_\Sigma |F|^2,~~~\mathcal E_2:=\int_\Sigma |\nabla^{(2)}\psi|^2,\\
\mathcal E_3&:=\int_\Sigma(4U|\psi|^2+\alpha|\psi|^4)
\end{align*}
The reason why we are decomposing the action in this way is because, in the topologically-trivial sector, all three terms can be simultaneously minimized: to minimize $\mathcal E_1,\mathcal E_2,$ and $\mathcal E_3$, it suffices to have, separately,
\begin{align*}
F(x)=0,~~~~\nabla^{(2)}\psi(x)=0,~~~~~|\psi(x)|^2=-\fr{2U}{\alpha}\tag{A.1}
\end{align*}
where the first condition only makes sense on a trivial line bundle. Since these equations are simultaneously satisfiable, the corresponding space of solutions modulo gauge transformations is exactly $\mathcal M_{GP}^\alpha(\Sigma)$.\\

\begin{center}
\includegraphics[scale=.45]{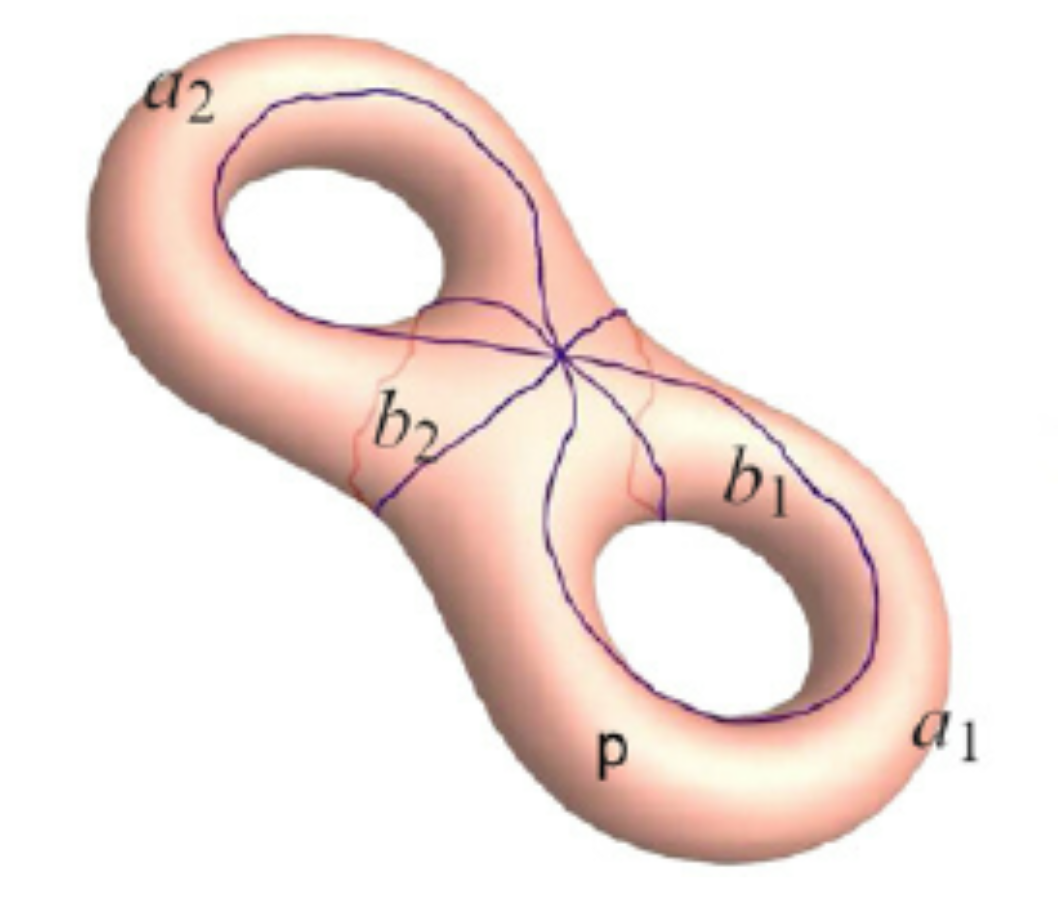}\\
{\small{\bf Figure 2}: {\it A superconducting groundstate on a surface $\Sigma$ (here $g=2$), in the absence of monopoles, is labeled by $2g$ bits, one for each generator of the fundamental group of the surface (Source: Google Images).}}
\end{center}

We now do as stated, that is, compute the solutions to the system of three minimization conditions (A.1). The first condition implies that $A$ is flat. The second condition implies that the value of the order parameter at any given point is the parallel transport of its value from anywhere else:
\[\psi(x)=\exp\left(i\int_x^y 2A\right) \psi(y)\]
Since $\psi$ is a global section, this implies that the holonomy of the connection $\nabla^{(2)}$ is always trivial inside the superconducting region. Since we have, for every closed loop $\gamma$,
\[\exp\left(i\int_x^y 2A\right)=\exp\left(i\int_x^y A\right)^2\]
We have that the holonomy of $A$ squares to one, and therefore lies in the subgroup $\ZZ_2\subset U(1)$. We now can state the following theorem:\\

\begin{thm}[Groundstate Subspace] We have an isomorphism of sets
\[\mathcal M_{GP}^\alpha(\Sigma)\simeq\mathcal M_\text{flat}(\Sigma,\,\ZZ_2) \]
where $\mathcal M_\text{flat}(\Sigma,\,\ZZ_2)$ denotes the holonomy $\pm 1$-subspace of the moduli space of flat connections on $\Sigma$ modulo gauge transformations.
\end{thm}
{\it Proof.}  Every groundstate of the superconductor, by the analysis above, corresponds to an equivalence class $[\ket{\psi,A}]$, where $A$ is a flat gauge field with holonomy $\pm 1$, and we have, for $\theta\in C^\infty(\Sigma)$,
\[\ket{\psi,A}\sim \ket{e^{i\theta}\psi,\,A-d\theta)}\] 
Therefore, we are essentially claiming that it suffices to represent each groundstate by the gauge-equivalence class of $A$, and forget about the order parameter.\\

This is because, to specify $\psi$, all we need is the value $\psi_0$ of the order parameter at a single point $x_0$ in the superconducting region; to obtain the value anywhere else, we parallel transport via $\nabla^{(2)}$:
\[\psi(x)=\exp\left(\int_{x_0}^x 2A\right)\psi_0.\]
Therefore, by specifying a value $\psi_0$ of the order parameter at a fixed point $x$, we can recover the groundstate which represents any given gauge field.\\

Once we choose such a value $\psi_0$ along with representatives $A_i$ from each equivalence class of flat gauge fields, we find that this defines a map of sets
\[A~~~\to ~~~(\exp\left(\int_{x_0}^x 2A\right)\,e^{i\theta}\psi_0,~~A)\tag{$A=A_i-d\theta$}\]
which factors onto a isomorphism $\mathcal M_{GP}^\alpha(\Sigma)\simeq\mathcal M_\text{flat}(\Sigma,\,\ZZ_2)$, upon modding-out by gauge transformations on both sides. $\square$

~\\

By our theorem, the groundstates of a superconductor are in one-to-one correspondence with the holonomy $\pm 1$-subspace of the moduli space of flat connections on $\Sigma$ modulo gauge transformations. By the universal coefficient theorem,
\begin{align*}
\mathcal M_\text{flat}(\Sigma,\ZZ_2)&\simeq \hom(\pi_1(\Sigma),\ZZ_2)\\
&\simeq \hom(H_1(\Sigma),\ZZ_2)\simeq H^1(\Sigma;\ZZ_2)
\end{align*}
and therefore we have an exact solution
\[\boxed{\mathcal M_{GP}^\alpha(\Sigma)=H^1(\Sigma;\mathbb Z_2)}\tag{A.2}\]
Note that, in deriving this formula, we did not have to assume that $\Sigma$ was two-dimensional. For the special case of a Riemann surface, however, we can go further: the rank-one cohomology of a Riemann surface with coefficients in $\ZZ_2$ is the direct product of $2g$ copies of $\ZZ_2$, where $g=0,1,2,\cdots$ is the genus. Therefore, 
\[|\mathcal M_{GP}^\alpha(\Sigma)|=2^{2g}\]
Therefore, we have, as desired, reproduced the results of \hyperlink{ref3}{[3]} demonstrating the one-to-one correspondence of groundstates of the BCS model to spin structures on $\Sigma$, without referring to the microscopic details of the BCS Hamiltonian.

\begin{table}[]
\centering
\begin{tabular}{|c|c|c|}
\hline
~$\Sigma$~& ~$H^1(\Sigma;\ZZ_2)$~ & ~~$|\mathcal M^\alpha_{GP}(\Sigma)|$~(GSD)~ \\ \hline
$S^1\times S^1$ & $\mathbb Z_2\times\mathbb Z_2$ & 4 \\ \hline
$\Sigma (g=2)$ & $\mathbb Z_2^4$ & 64 \\ \hline
$\Sigma (g=3)$ & $\mathbb Z_2^6$ & 256 \\ \hline
\end{tabular}
\caption{The groundstate degeneracy of a monopole-free BCS model on a surface $\Sigma$ exhibits a marked dependence on the topology of $\Sigma$.}
\label{my-label}
\end{table}

\section{The BCS Groundstate Sector with Monopoles,  at $\alpha=1/4$ }
Now we allow ourselves to be in a topologically nontrivial sector, and allow $n\neq 0$. The calculation of the moduli space of groundstates of the BCS model will be vastly more difficult in this case, and so we focus on the special case $\alpha=1/4$, where the GP energy functional coincides with the Yang-Mills-Higgs action functional. Roughly,
\[\mathcal E^{1/4}_{GP}\propto S_{YMH}+\text{const.}\]
This will allow us to efficiently extract a large subspace of $\mathcal M^{1/4}_{GP}(\Sigma)$, i.e., a large class of superconducting groundstates, in spite of the complications caused by the net magnetic flux through the surface.

\subsection{The Yang-Mills-Higgs Moduli space}
We now introduce the {\bf Yang-Mills-Higgs action}, at level $\tau$, on a complex line bundle $\mathcal L\to \Sigma$, by the following expression:
\begin{align*}
S_{YMH}^\tau(\psi,A)=S_{YM}(A)+\int_\Sigma |\nabla \psi|^2+\fr{1}{4}(\tau-|\psi|^2)^2.
\end{align*}
The moduli space of solutions modulo gauge transformations to the corresponding Yang-Mills-Higgs equations in the topological sector $c_1(\mathcal L)=n$
was computed by Bradlow in 1990 \hyperlink{ref4}{[4]} to be, in the limit $\text{Vol}(\Sigma)\gg \tau$, isomorphic to the $n$-fold symmetric product of the surface with itself:
\[\mathcal M_{YMH}^\tau(\Sigma)\simeq S^n\Sigma\]
Therefore, if we treated the Yang-Mills-Higgs action as an energy density $\mathcal E_{YMH}^\tau$ for a quantum many-body Hamiltonian, i.e., identical to the situation with $\mathcal E_{GP}^\alpha$, then the groundstate sector for this Hamiltonian would be isomorphic to the $n$-particle component of the bosonic Fock space on $\Sigma$, i.e. each groundstate $\ket{\Omega_{YMH}}$ would admit a labelling by vortex locations:\[\ket{\Omega_{YMH}} = \ket{\{z_1,\cdots z_n\}},~~~z_i\in \Sigma.\]

\subsection{Injecting the Yang-Mills-Higgs moduli space into the space of superconducting groundstates}
We now take advantage of the miraculous coincidence  between the Yang-Mills-Higgs functional at $\tau=-8U$ and the superconducting energy functional at $\alpha=1/4$ to extract crucial information about the superconducting groundstates in the presence of monopoles. The relation is most succinctly stated as
\begin{align*}
S_{YMH}^{-8U}(\psi, A)&\propto \mathcal E_{GP}^{1/4}(\psi,A/2)+\text{const}.
\end{align*}
where the constant of proportionality is positive. Therefore, if we take any equivalence class $[\ket{\psi,A}]$ in the Yang-Mills-Higgs moduli space, then, automatically, $\ket{\psi,A/2}$ is an equivalence class of superconducting groundstates:
\[[\ket{\psi,A/2}]\in \mathcal M_{GP}^{1/4}(\Sigma).\tag{A.3}\]
Furthermore, if $[\ket{\psi',A'}]\neq [\ket{\psi,A}]\in \mathcal M^{-8U}_{YMH}(\Sigma)$ are distinct in the Yang-Mills-Higgs moduli space, then the corresponding superconducting groundstates constructed via (A.3) are also distinct in the Gross-Pitaevski moduli space:
\[[\ket{\psi',A'/2}]\neq [\ket{\psi,A/2}]\in  \mathcal M_{GP}^{1/4}(\Sigma).\]
To see this, suppose to the contrary that
\[[\ket{\psi',A'/2}]= [\ket{\psi,A/2}]\in  \mathcal M_{GP}^{1/4}(\Sigma).\]
Then, in particular, this means that there exists a gauge transformation $g$, such that, in a local trivialization
\[\psi'=e^{i2\theta}\psi,~~~~A'/2=A/2-d\theta.\tag{$g=e^{i\theta}$}\]
Therefore, letting $h$ be the gauge transformation $g$ composed with itself twice, then, in that same trivialization, 
\begin{align*}
\psi'=e^{i\phi}\psi,~~~~A'=A-d\phi.\tag{$h=e^{i\phi}$}
\end{align*}
Therefore, since this analysis holds in each local trivialization, we have $[\ket{\psi',A'}]=[\ket{\psi,A}]\in \mathcal M^{-8U}_{YMH}(\Sigma)$, contradicting our original assumption that the two classes were unequal in the Yang-Mills-Higgs moduli space. Therefore, in total, we have constructed an injection of moduli spaces
\begin{align*}
\mathcal M_{YMH}^{-8U}(\Sigma)&\hookrightarrow \mathcal M_{GP}^{1/4}(\Sigma),
\end{align*}
sending $[|\psi,A\rangle]\mapsto [|\psi,A/2\rangle]
$, that allows us to view $\mathcal M_{YMH}^{-8U}(\Sigma)$ as a subset of $\mathcal M_{GP}^{1/4}(\Sigma)$. This gives us a lower-bound on the number of superconducting groundstates on $\Sigma$ in this nontrivial setting, where it is relatively difficult to obtain information.
\begin{center}
\includegraphics[scale=.4]{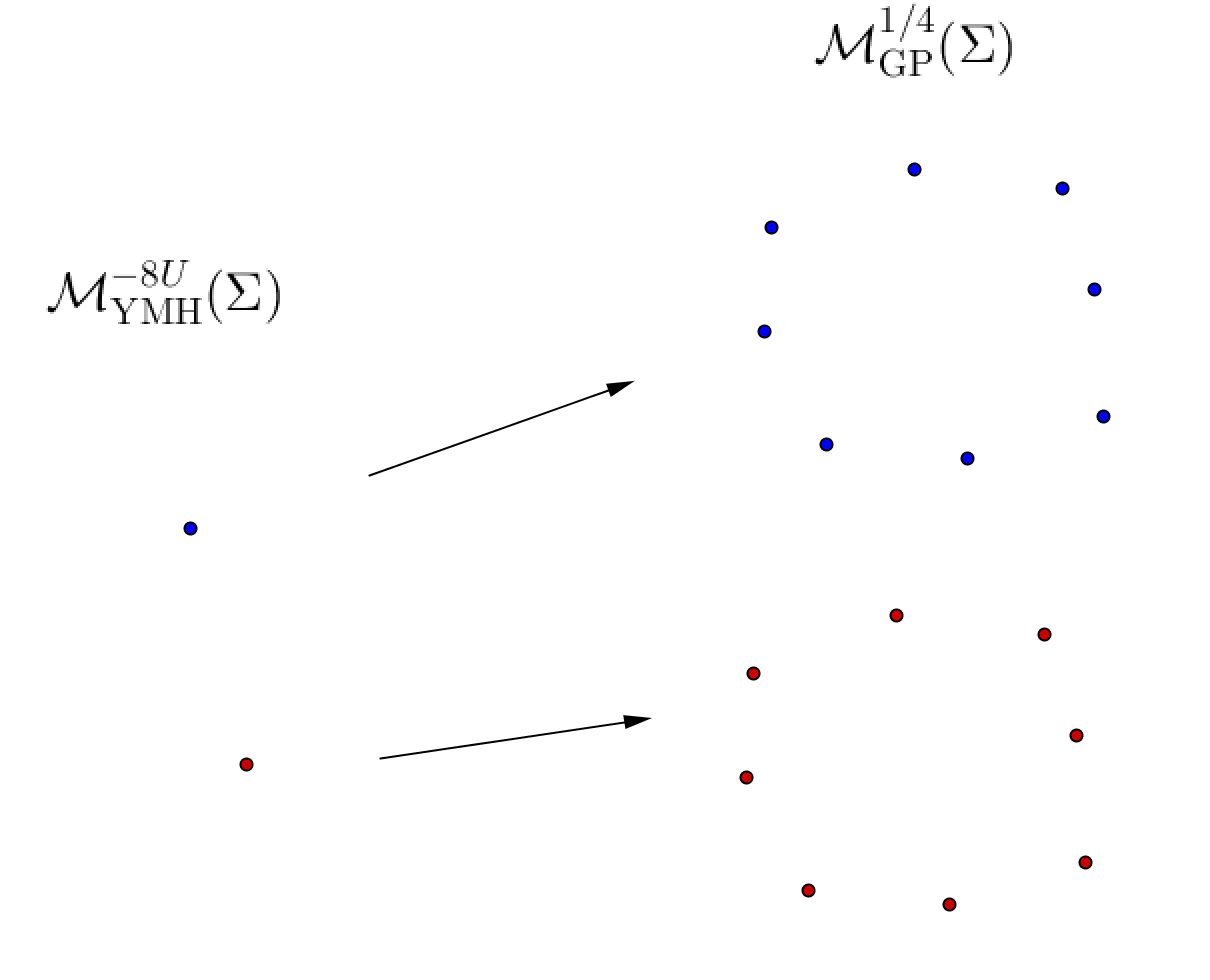}
{\small {\bf Figure 3}. {\it  The refined construction: each equivalence class of solutions to the Yang-Mills-Higgs equations splits into a distinct family of gauge-inequivalent solutions to the $\alpha=1/4$ GP equations.}}
\end{center}

\noindent In particular, the set of superconducting groundstates in the image of this injection, by Bradlow's calculation earlier, is isomorphic to the $n$-fold symmetric product of the surface with itself:
\begin{align*}
S^n\Sigma\subset \mathcal M_{GP}^{1/4}(\Sigma)
\end{align*}
Therefore, identical to the situation earlier, each superconducting groundstate $\ket{\Omega_\text{BCS}}$ in this subset admits a labelling by vortex locations:
\begin{align*}
\ket{\Omega_{BCS}}=\ket{\{z_1,\cdots z_n\}},~~~~z_i\in \Sigma.\\
\end{align*}

\subsection{Refining the Lower-Bound}
Having sketched a basic construction, namely the injection of the Yang-Mills-Higgs moduli space into the Gross-Pitaevskii moduli space at $\alpha=1/4$, we can refine this construction in a way that will increase the lower bound by a sizable topological factor. In particular, for each class $\ket{\psi,A}\in \mathcal M_{YMH}^{-8U}(\Sigma)$, we actually can construct up to $2^{2g}$ distinct superconducting groundstates
\[\left[\ket{\psi,A(\sigma_1,\cdots, \sigma_{2g})/2}\right]\in \mathcal M_{GP}^{1/4}(\Sigma),~~~~\sigma_i\in \ZZ_2\]
defined by the integral equations
\begin{align*}
\exp\left(i\int_{\gamma_j} A(\sigma_1,\cdots \sigma_{2g})/2\right)=\sigma_j,
\end{align*}
where $\gamma_1,\cdots,\gamma_{2g}$ are the fundamental cycles of the surface. These are manifestly gauge inequivalent, because holonomies are invariant under gauge transformations. Therefore,  each solution of the Yang-Mills-Higgs equations splits into a family of $2^{2g}$ physically-distinct superconducting groundstates. This is the topological fractionalization phenomenon detailed in {\it Fractionalization, Topological Order, and Quasiparticle Statistics}. Therefore, the initial estimate as to the number of superconducting groundstates at $\alpha=1/4$ can be multiplied by the rank of the holonomy $\pm 1$ subspace of the moduli space of flat connections on $\Sigma$:
\[ \boxed{\mathcal M_\text{flat}(\Sigma,\ZZ_2)\times S^n\Sigma\,\subset \mathcal M^{1/4}_{GP}(\Sigma)}\]
Unpacking the physics, this class of superconducting groundstates is labelled by a combination of both continuous quantum numbers, corresponding to vortex locations, and discrete quantum numbers, corresponding to holonomy of $A$ around the generators $\gamma_1,\cdots , \gamma_{2g}$ of the fundamental group of $\Sigma$:
\begin{align*}
\ket{\Omega_{BCS}}&=\ket{\{z_1,\cdots z_n\},\sigma_1,\cdots, \sigma_{2g}},~~~~\\
&~~~~~~~~~~~~~~~~~~~~~~~~~~z_i\in \Sigma, ~~\sigma_i\in \ZZ_2.
\end{align*}
Therefore, we have found a continuum of superconducting groundstates, labelled by a mixture of both discrete and continuous quantum numbers.

\subsection{Example: Groundstate Sector of BCS Theory at $\alpha=1/4$ on the Hopf Bundle $S^3$}
We now focus on a concrete example, to illuminate the relevant characteristics of the classification program we have carried out for two-dimensional BCS theory. Therefore, we will solve the $\alpha=1/4$ GP equations on the bundle $\mathcal L_\text{Hopf}\to S^2$, the associated line bundle to the Hopf bundle $S^3\to S^2$.\\
\begin{center}
\includegraphics[scale=.5]{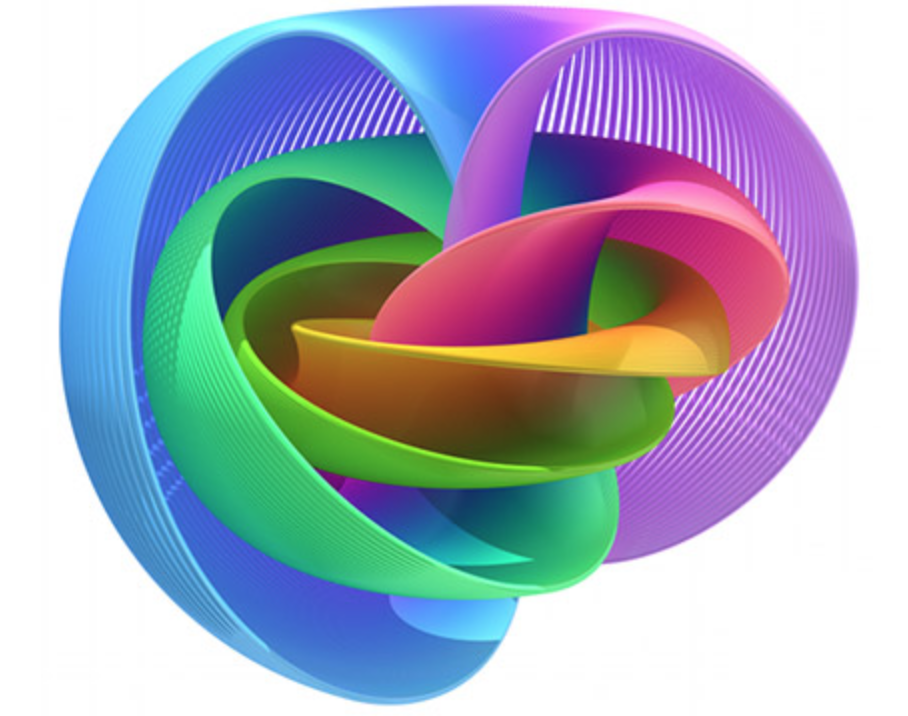}\\
{\it {\bf Figure 4}. The Hopf Fibration (Source: Google Images).}\\
\end{center}

Since the associated bundle $\mathcal L_\text{Hopf}$ has Chern number $c_1(\mathcal L_\text{Hopf})=1$, and since $\mathcal M_\text{flat}(S^2,\ZZ_2)=0$,  our technique extracts a class of groundstates
\[\boxed{S^2\subset \mathcal M_{GP}^{1/4}(S^2)}\]
in one-to-one correspondence with the surface $S^2$ itself. Physically, this makes sense: since there is only one vortex, each superconducting groundstate is labelled by the location $z\in S^2$ of that vortex. Furthermore, all loops on a two-dimensional sphere can be contracted to a point, and so a flat gauge field has trivial holomomy; so there is no additional ground state degeneracy (GSD) that we can deduce from fractionalization arguments.\\

\section{Summary and Outlook}
The physics of superconductivity has been incredibly successful in inspiring revolutionary technologies such as SQUIDs, superconducting qubits, MRI, and more. In this paper, we examined the topological groundstate degeneracy (GSD) of two-dimensional superconductors with non-zero net magnetic flux. In the already well-known case $n=0$, we reproduced the results of Read and Green in \hyperlink{ref3}{[3]} by explicitly identifying superconducting groundstates on a surface $\Sigma$ with the set of spin structures on that surface:
\[\mathcal M^{\alpha}_{GP}(X)\simeq H^1(X;\ZZ_2),~~~~\mathcal L\simeq \underline{\CC}\,,~~~~\dim X=1,2,\cdots\]
In dimension two and $n\neq 0$, we witnessed new physics, including a continuum of superconducting groundstates, parametrized both by continuous vortex locations as well as discrete flux values around non-contractible loops on the surface. This was proven by our identification of a large class of solutions to the  Gross-Pitaevskii equations at $\alpha=1/4$:
\[H^1(X;\ZZ_2)\times S^n\Sigma\,\subset \mathcal M^{1/4}_{GP}(\Sigma),~~~~~~\dim \Sigma=2\]
Once again, study of superconductivity over the past century has inspired new technologies, and we hope that, by pushing the boundaries of superconducting physics, the results will be no different, inspiring new technological paradigms.\\

\section{Acknowledgements}
I would like to acknowledge my thesis advisor Clifford Taubes for being an incredible mentor during my senior year  at Harvard. I am also grateful for many fruitful discussions with Bertrand Halperin, from whom,  throughout my last three semesters of college, I learned much condensed matter theory. Finally, I would like to thank Harvard junior Jake McNamara for helping me understand the relevant aspects of Spin structures. Finally, I also am indebted to Peter Kronheimer at Harvard University, and Andre Petukhov and Sergey Knysh at NASA QuAIL for providing useful feedback on this work.\\

\newpage

\section{References}

\hypertarget{ref1}{}
\begin{enumerate}
\item[\hypertarget{ref2}{[1]}] Christian Hainzl, Benjamin Schlein, {\it Dynamics of Bose-Einstein condensates of fermion pairs in the low density limit of BCS theory}, (2001), \url{https://arxiv.org/abs/1203.2811}.

\item[\hypertarget{ref3}{[2]}]  Laszlo Erdos, Benjamin Schlein, Horng-Tzer Yau, {\it Derivation of the Gross-Pitaevskii equation for the dynamics of Bose-Einstein condensate}, (2010), Annals of Mathematics.

\item[\hypertarget{ref4}{[3]}] N. Read, Dmitry Green, {\it Paired states of fermions in two dimensions with breaking of parity and time-reversal symmetries, and the fractional quantum Hall effect}, Phys. Rev. B, (2000), \url{https://arxiv.org/abs/cond-mat/9906453}

\item[\hypertarget{ref5}{[4]}] Bradlow, {\it Vortices in holomorphic line bundles over closed Kahler manifolds}, Communications in Mathematical Physics, (1990), \url{http://projecteuclid.org/euclid.cmp/1104201917}

\item[\hypertarget{ref6}{[5]}] Jaffe, A., Taubes, C., {\it Vortices and Monopoles, Structure of Static Gauge Theories}, Progress in Physics 2, Boston-Basel-Stuttgart, Birkhauser Verlag (1980), 287 S., DM 30,-. ISBN 3-7643-3025-2

\item[\hypertarget{ref7}{[6]}] Masaki Oshikawa, T. Senthil, {\it Fractionalization, topological order, and quasiparticle statistics}, Phys. Rev. Lett. 96, 060601, (2006), \url{https://arxiv.org/abs/cond-mat/0506008}

\end{enumerate}

\end{document}